\documentclass[twocolumn,aps,nofootinbib]{revtex4-1}
\usepackage{graphicx}
\usepackage{booktabs}
\usepackage{xcolor}
\usepackage{hyperref}
\usepackage{multirow}
\usepackage{setspace}
\usepackage{amsmath, amssymb, amsthm}
\usepackage{cleveref}
\usepackage{physics}
\newcommand{\comment}[1]{}

\newcommand{\etal}{\textit{et al.}}

\begin{document}
\author{Alexander F. Siegenfeld$^{1,2}$} 
\email{Corresponding author.  Email: asiegenf@mit.edu}
\author{Yaneer Bar-Yam$^2$}
\affiliation{$^1$Department of Physics, Massachusetts Institute of Technology,  Cambridge, MA}
\affiliation{$^2$New England Complex Systems Institute, Cambridge, MA}
\title{An Introduction to Complex Systems Science and its Applications}

\begin{abstract}
The standard assumptions that underlie many conceptual and quantitative frameworks do not hold for many complex physical, biological, and social systems.  Complex systems science clarifies when and why such assumptions fail and provides alternative frameworks for understanding the properties of complex systems.  
This review introduces some of the basic principles of complex systems science, including complexity profiles, the tradeoff between efficiency and adaptability, the necessity of matching the complexity of systems to that of their environments, multi-scale analysis, and evolutionary processes.  Our focus is on the general properties of systems as opposed to the modeling of specific dynamics; rather than provide a comprehensive review, we pedagogically describe a conceptual and analytic approach for understanding and interacting with the complex systems of our world.   With the exception of a few footnotes, this paper assumes only a high school mathematical and scientific background, so that it may be accessible to academics in all fields, decision-makers in industry, government, and philanthropy, and anyone who is interested in systems and society.   
\end{abstract}

\maketitle

\section{Introduction}
How can we scientifically approach the study of complex systems---physical, biological, and social?  Empirical studies, while useful, are by themselves insufficient, since all experiments require a theoretical framework in which they can be interpreted.  While many such frameworks exist for understanding particular components or aspects of systems, the standard assumptions that underlie most quantitative studies often do not hold for systems as a whole, resulting in a mischaracterization of the causes and consequences of large-scale behavior. 

This paper provides an introduction to complex systems science, demonstrating a few of its applications and its capacity to help us make more effective decisions in the complex systems of our world.  It focuses on some general properties of complex systems, rather than on the modeling of specific dynamics as in the subfields of dynamical systems, agent-based modeling and cellular automata, network science, and chaos theory.
Section~\ref{sec:basic} introduces key concepts, including complexity profiles, the tradeoff between efficiency and adaptability, and the necessity of matching the complexity of systems to that of their environments. Section~\ref{sec:analyze} considers the analysis of complex systems, attending to the oft-neglected question of when standard assumptions do and---more importantly---do not apply.  Section~\ref{sec:decision}  discusses principles for effectively intervening in complex systems given that their full descriptions are often beyond the limits of human comprehension.  Section~\ref{sec:further} provides further reading.  Section~\ref{sec:conclusion} concludes.

\section{Basic Principles of Complex Systems Science}
\label{sec:basic}
\subsection{Why complex systems science?}
\label{sec:why}
\begin{figure}
\begin{center}
\includegraphics[width=.5\textwidth]{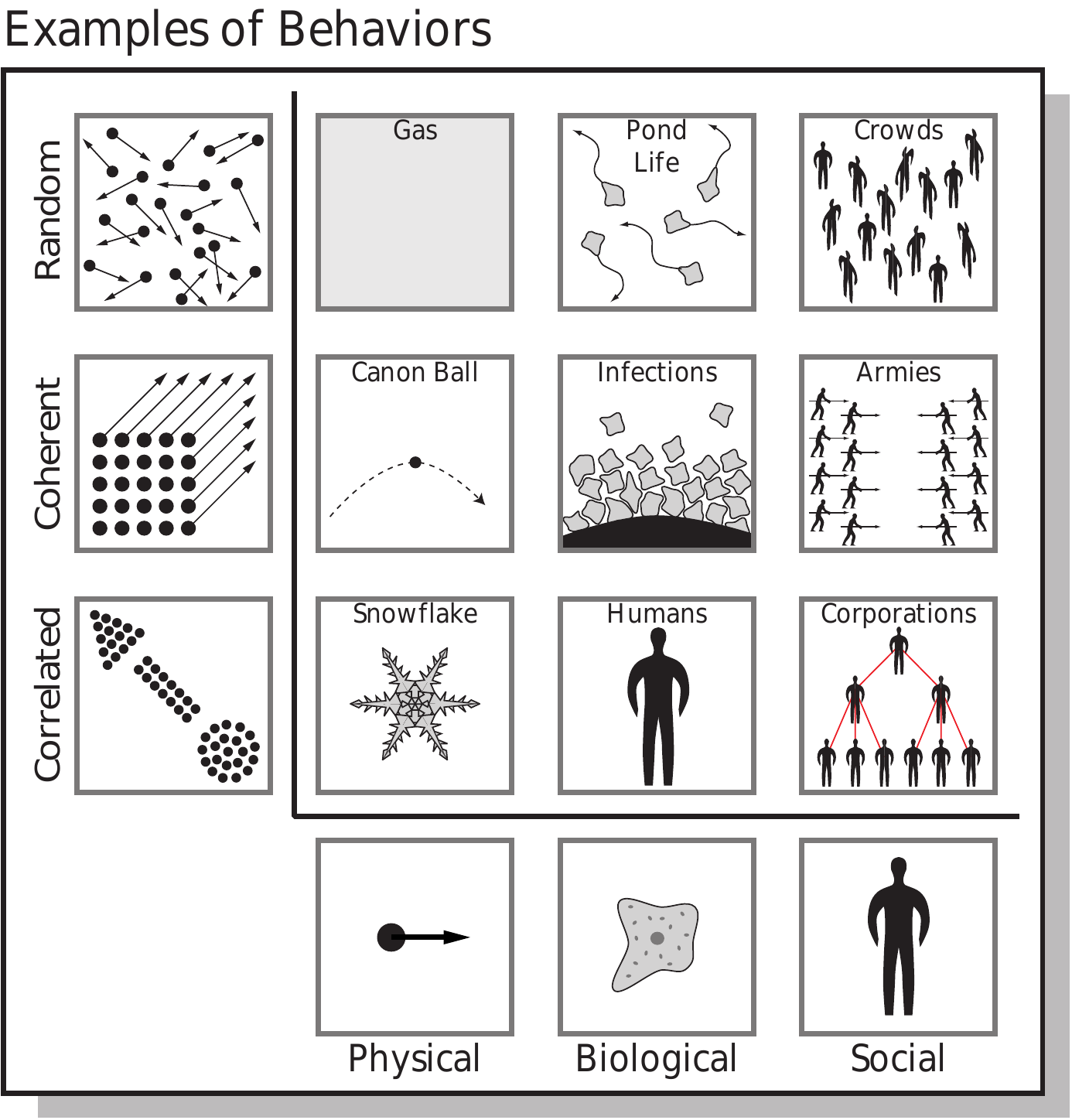}
\caption{From ref.~\cite{behaviors}.  Each column contains three examples of systems consisting of the same components (from left to right: molecules, cells, people) but with different relations between them.  Each row contains systems representing a certain kind of relationship between components.  For random systems, the behavior of each component is independent from the behavior of all other components.  For coherent systems, all components exhibit the same behavior; for example, the behavior (location, orientation, and velocity) of one part of the cannonball completely determines the behavior of the other parts.  Correlated systems lie between these two extremes, such that the behaviors of the system's components do depend on one another, but not so strongly that every component acts in the same way; for example, the shape of one part of a snowflake is correlated with but does not completely determine the shape of the other parts.  (Implicit in these descriptions is the necessity of specifying the set of behaviors under consideration, as discussed in \cref{sec:complexity}.)}
\label{fig:behaviors}
\end{center}
\end{figure}

Complex systems science considers systems with many components.  These systems could be physical, biological, or social.  Given this diversity of systems, it may seem strange to study them all under one framework.  But while most scientific disciplines tend to focus on the components themselves, complex systems science focuses on how the components within a system are related to one another~\cite{textbook}.  For instance, while most academic disciplines would group the systems in \cref{fig:behaviors} by column, complex systems science groups them by row.

Systems may differ from each other not because of differences in their parts but because of differences in how these parts depend on and affect one another.   For example, steam and ice are composed of identical water molecules but, due to differences in the interactions between the molecules, have very different properties.  Conversely, all gasses share many behaviors in common despite differences in their constituent molecules.  The same holds for solids and liquids.   The behaviors that distinguish solids from liquids from gasses are examples of \textit{emergence}: they cannot be determined from a system's parts individually.  Fluid turbulence, as one might observe in a flowing river, is an example of how the relationships between parts can give rise to emergent large-scale behaviors and patterns that are self-organized, meaning that they arise not from some external or centralized control but rather autonomously from the interactions between the system components~\cite{anderson1972more,kauffman1993origins,eigen1971selforganization,cross1993pattern,haken2006information}.  Other examples of self-organized behaviors include the spontaneous formation of conversation groups at a party, the allocation of goods in a decentralized economy, the evolution of ecosystems, and the flocking of birds.   Such large-scale behaviors and patterns cannot be determined by examining each system part in isolation.  By instead considering general properties of systems as wholes, complex systems science provides an interdisciplinary scientific framework that allows for the discovery of new ideas, applications, and connections.

A full description of all the small-scale details of even relatively simple systems is impossible; therefore sound analyses must describe only those properties of systems that do not depend on all these details.
That such properties exist is due to \textit{universality}, a phenomenon that will be discussed in \cref{sec:analyze}.  Statistical physics provides an underlying insight that allows for the discovery of such properties: namely, that while attempting to characterize the behavior of a particular state of a system (e.g. a gas) may be entirely intractable, characterizing the set of all possible states of the system may not only be tractable but may also provide us with a model of the relevant information (e.g. the pressure, temperature, density, compressibility, etc.).  In other words, taking a step back and considering the \textit{space of possible behaviors} provides a powerful analytical lens that can be applied not only to physical systems but also to biological and social ones.

\subsection{What is complexity?}
\label{sec:complexity}
We define the complexity of a behavior as equal to the length of its description.  The length of a description of a particular system's behavior depends on the number of possible behaviors that system could exhibit~\cite{cover2012elements}.  For example, a light bulb that has two possible states---either on or off---can be described by a single bit: 0 or 1.  Two bits can describe four different behaviors (00, 01, 10, or 11),  three bits can describe eight behaviors, and so on.  Mathematically, we can write $C=\log_2 N$, where $C$ is the complexity of a system and $N$ is its number of possible behaviors,\footnote
{Technically, $\log_2 N$ is actually an upper bound for the system's complexity since if some behaviors are more likely than others, the average length of the system's description can be reduced by using shorter descriptions for the more common behaviors and longer descriptions for the less common ones.  (Lossless compression algorithms rely on this logic.)}
but for our purposes here, it is sufficient to state that the greater the number of possible behaviors, the greater the complexity.  

It is important to note that one must carefully define the space of possible behaviors.  For instance, if we are interested in a light bulb already in a socket, the light bulb has two possible behaviors, as above, but if we are instead interested in the complexity of building a light bulb, the space of possible behaviors might include all of the ways in which its parts could be arranged.  As another example, consider programming a computer to correctly answer a multiple-choice question with four choices.  At first glance, this task is very simple: since there are four possible behaviors, only two bits are required.  Nonetheless, we have the sense that programming a computer to score perfectly on a multiple-choice test would be quite difficult.  This apparent paradox is resolved, however, when we recognize that such a task is difficult only because we do not \textit{a priori} know what questions will be on the test, and thus the true task is to be able to correctly answer \textit{any} multiple-choice question.  This task is quite complex, given the large number of possible ways the program could respond to a string of arbitrary multiple-choice questions.

\subsection{Complexity and scale}
\label{sec:cx}
Consider a human, and then consider a  gas containing the very same molecules that are in the human but in no particular arrangement.  Which system is more complex?  The gas possesses a greater number of possible arrangements of the molecules (i.e. has more entropy, or disorder), and thus would take longer to describe at a microscopic level.  However, when we think of a complex system, we think of the behaviors arising from the ordered arrangement of molecules in a human, not the behaviors arising from the maximally disordered arrangement of molecules in a gas.  
It therefore may be tempting to conclude that complex systems are those with reduced disorder.  But the systems with the least disorder are those in which all components exhibit the same behavior (coherent systems in \cref{fig:behaviors}), and such behavior is easy to describe and thus not intuitively complex.  

To resolve this apparent paradox, we must consider that the length of a system's description depends on the level of detail used to describe it.  Thus, complexity depends on scale.  On a microscopic scale, it really is more difficult to describe the positions and velocities of all the molecules of the gas than it is to do the same for all the molecules of the human.  But at the scale of human perception, the behaviors of a gas are determined by its temperature and pressure, while the behaviors of a human remain quite complex.  Entropy corresponds to the amount of complexity at the smallest scale, but characterizing a system requires understanding its complexity across multiple scales.  A system's \textit{complexity profile} is a plot of the system's complexity as a function of scale~\cite{allen2017multiscale}.  In the examples below, scale will be taken to be length, but fundamentally, the scale of a behavior is equal to the number of coordinated components involved in the behavior,
for which physical length is a proxy.  A gas is very simple at the scale of human perception because at this scale, only behaviors involving trillions of molecules are relevant, and there are relatively few distinguishable behaviors of a gas involving so many molecules.

\begin{figure}
\begin{center}
\includegraphics[width=.5\textwidth]{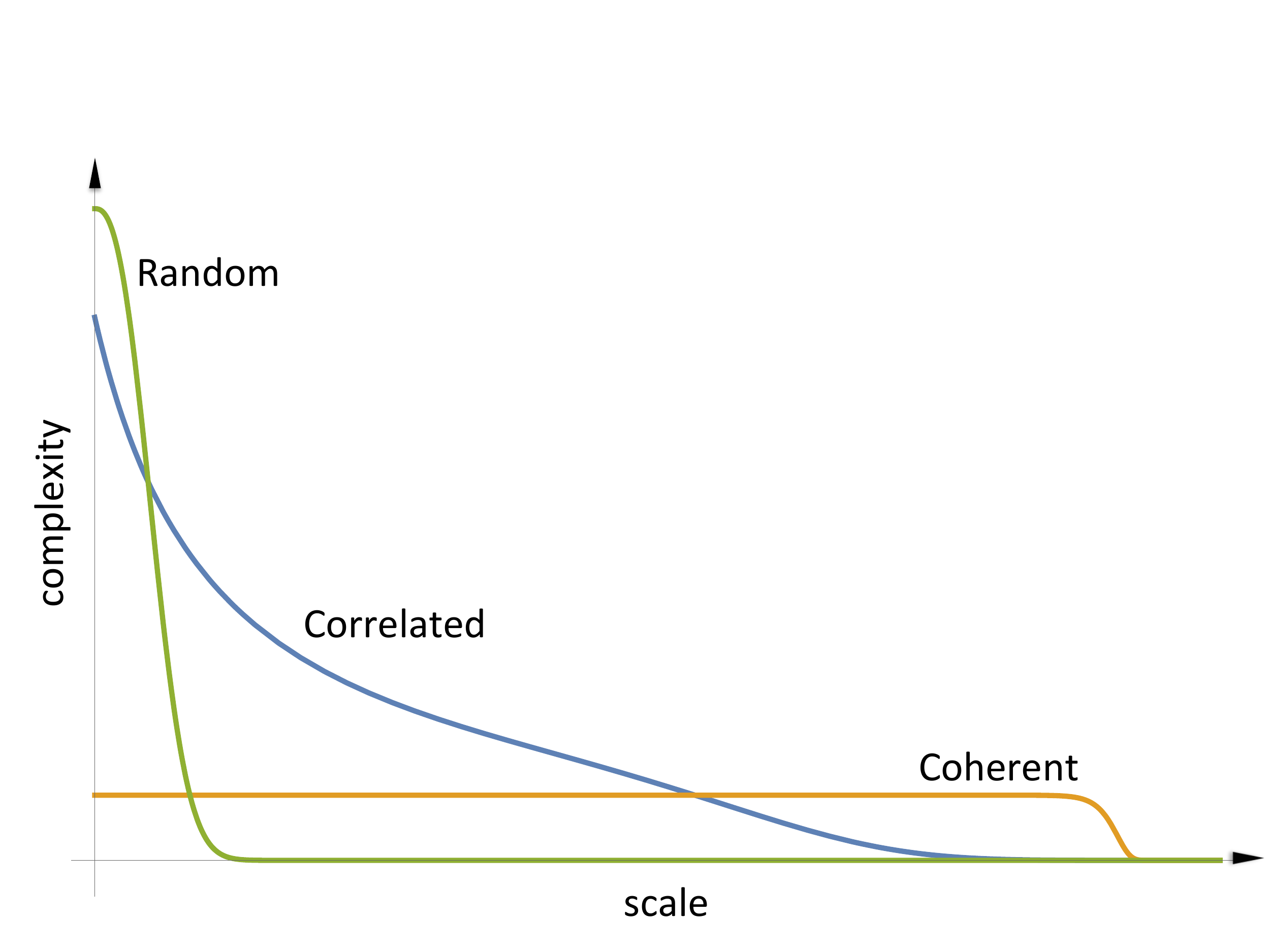}
\caption{Representative complexity profiles for random, coherent, and correlated systems (see \cref{fig:behaviors}).  Any given system may have aspects of each at various scales.}
\label{fig:profile}
\end{center}
\end{figure}

As shown in \cref{fig:profile}, random, coherent, and correlated systems (see \cref{fig:behaviors}) have qualitatively different complexity profiles.  Random systems have the most complexity at the smallest scale (finest granularity/most detail), but the amount of complexity rapidly drops off as the scale is increased and the random behaviors of the individual components are averaged out.  A coherent system has the same amount of complexity at small scales as it does at larger scales because describing the overall behavior of the system (e.g. the position and velocity of a cannonball) also describes the behavior of all the components (e.g. the positions and velocities of all the atoms).  Note that complexity tends to increase (or remain the same) as the scale decreases, since looking at a system in more detail (while still including the whole system in the description) tends to yield more information.
For a correlated system, various behaviors occur at various scales, and so the complexity gradually increases as one examines the system in greater and greater detail.  For instance, from very far away a human, being barely visible, has very little complexity.  As the level of detail is gradually increased, the description will first include the overall position and velocity of the human, and then the positions and velocities of each limb, followed by the movement of hands, fingers, facial expressions, as well as words that the human may be saying.  Continuing to greater levels of detail, the organs and then tissues and  patterns within the human brain become relevant, and eventually so do the individual cells.  At scales smaller than that of a cell, complexity further increases as one sees organelles (cellular substructures), followed by large molecules such as proteins and DNA, and then eventually smaller molecules and individual atoms.  At each level, the length of the description grows longer.  This incredible multi-scale structure with gradually increasing complexity is a defining characteristic of complex systems.

\subsection{Tradeoffs between complexity and scale}
\label{sec:tradeoffs}
The intuition that complex systems require order is not unfounded: for there to be complexity at larger scales, there must be behaviors involving the coordination of many smaller-scale components.  This coordination suppresses complexity at smaller scales because the behaviors of the smaller-scale components are now limited by the interdependencies between them.  The tension between small-scale and large-scale complexity can be made precise: given a fixed set of components with a fixed set of potential individual behaviors, the area under the complexity profile will be constant, regardless of the interdependencies (or lack thereof) between the components.\footnote
{Formally, the sum of a system's complexity at each scale (i.e. the area under its complexity profile) will equal the total complexity of its components, i.e. the sum of each individual component's complexity~\cite{allen2017multiscale}.  The complexity of an individual component is related to the number of distinct behaviors of that component, as described in \cref{sec:complexity}.}  
Thus, for any system, there is a fundamental tradeoff between the number of behaviors a system can have and the scale of those behaviors.

\begin{figure}
\begin{center}
\includegraphics[width=.5\textwidth]{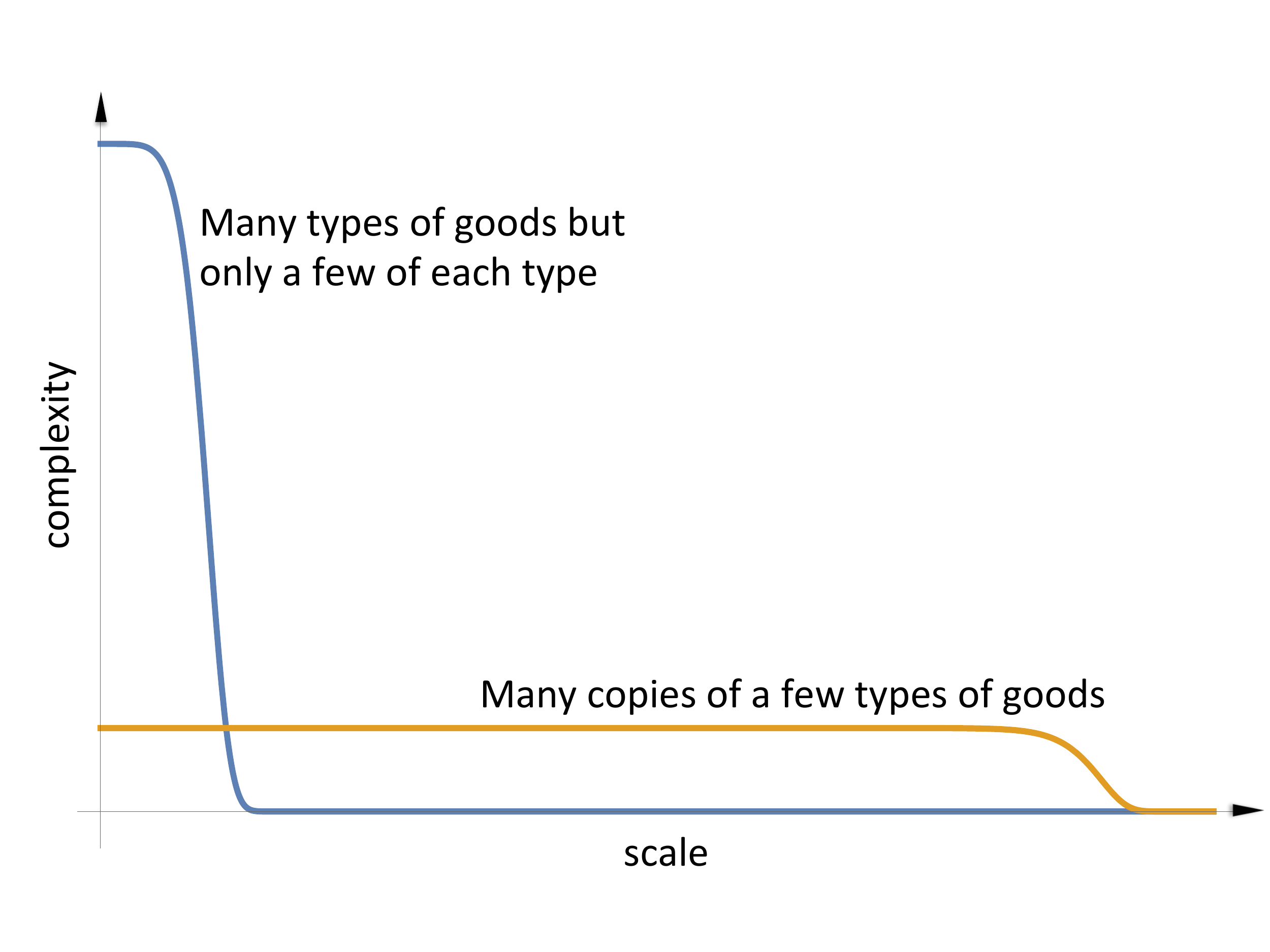}
\caption{The complexity profile of a factory that can produce a large number of copies of a few types of goods, and the complexity profile of a factory that can produce many types of goods but not in large numbers.  The number of copies of a good produced is a proxy for scale since, given a fixed technology, mass production requires larger-scale coordinated action in the factory (e.g. an assembly line), and the number of different types of goods that can be produced at a given scale is a proxy for the number of different possible behaviors of the factory---and thus its complexity---at that scale.}
\label{fig:factory}
\end{center}
\end{figure}

For instance, consider a factory consisting of many workers~\cite{behaviors}.  The output of the factory can be characterized using a complexity profile (\cref{fig:factory}).  The number of different types of goods that the factory can produce at a given scale is a proxy for the factory's complexity at that scale, with the number of copies of the same type of good that the factory can produce in a given amount of time being a proxy for scale.  The fundamental tradeoff is evident in the fact that if the factory wants to be able to churn out many copies of a single type of good in a short amount of time, it will have to coordinate all of its workers (perhaps having them work on an assembly line), thereby reducing their individual freedom to make many different kinds of goods.  The factory's production would then have low complexity but at a large scale (e.g. churning out many identical Model-T Fords---``Any customer can have a car painted any color that he wants so long as it is black").  On the other hand, if the factory's employees work independently, they will be able to create many different types of products, but none at scale.  
Of course, a factory may be able to increase both the complexity and scale of its production by adding new machinery or more workers; the precise tradeoff between complexity and scale applies only when considering a fixed set of components with a fixed set of individual behaviors.\footnote
{A subtle point to be made here is that introducing interactions between two parts of a system may in some cases increase the set of relevant individual behaviors of each part, thereby increasing the total area under the complexity profile.  For example, if two people enter into communication with each other, the communication itself (e.g. speech) may now be a relevant behavior of each individual person that was not there before.}

A corollary of the tradeoff between complexity and scale is the tradeoff between adaptability and efficiency~\cite{ulanowicz2002balance,korhonen2008beyond,ulanowicz2009dual,weigelt2012performance,pizzol2013network,panyam2019bio}.  Adaptability arises when there are many possible actions happening in parallel that are mostly independent from one another, i.e. when the system has high complexity.  Efficiency, on the other hand, arises when many parts of a system are all working in concert, so that the system can perform the task for which it was designed at the largest possible scale.  Due to the tradeoff between complexity and scale, a system with more adaptability will have a complexity profile with greater complexity but predominantly at smaller scales, while a system with more efficiency will have a complexity profile with lower complexity but extending to larger scales.  Thus, a very efficient system will, due to its necessarily lower complexity, not be as adaptable to unforeseen variations within itself or its environment, while a very adaptable system, designed to handle all sorts of shocks, will necessarily have to sacrifice some larger-scale behaviors.     The Soviets thought they could have their cake and eat it, too: they originally believed that their economy would outperform capitalist ones because capitalist economies have so much waste related to multiple businesses competing to do the same thing~\cite[Chapter~16]{mtw}.  It would be far more efficient to coordinate all economic production.  But in creating such large-scale economic structures, lower-scale complexity was sacrificed, resulting in a non-adaptive system.  (Improperly regulated capitalist systems may also sacrifice redundancy and adaptability for efficiency, resulting in, for instance, excessive concentrations of market power, harmful feedback loops, and herd-like behaviors~\cite{de1990positive,lux1999scaling,lietaer2010our,Bouchaud2013,harmon2015anticipating}.)  

Due to the tradeoff between complexity and scale, any mechanism that creates larger-scale complexity---whether market or government or otherwise---will necessarily reduce individual complexity.  This is not to say that larger-scale complexity is always harmful; it is often worth trading some individual-level freedoms for larger-scale cooperation.  When, then, is complexity at a particular scale desirable?
 
\subsection{Why be complex?}
\label{sec:ashby}
A determination of when complexity is desirable is provided by the \textit{Law of Requisite Variety}~\cite{ashby1991requisite}: To be effective, a system must be at least as complex as the environmental behaviors to which it must differentially react.    If a system must be able to provide a different response to each of 100 environmental possibilities and the system has only 10 possible actions, the system will not be effective.  At the very least the system would need 100 possible actions, one for each scenario it could encounter.  (The above condition is necessary but of course not sufficient; a system with sufficiently many actions may still not take the right actions in the right circumstances.)  Note that the environment to which a system must react is itself also a system and will sometimes be referred to as such. 

\begin{figure}
\begin{center}
\includegraphics[width=.5\textwidth]{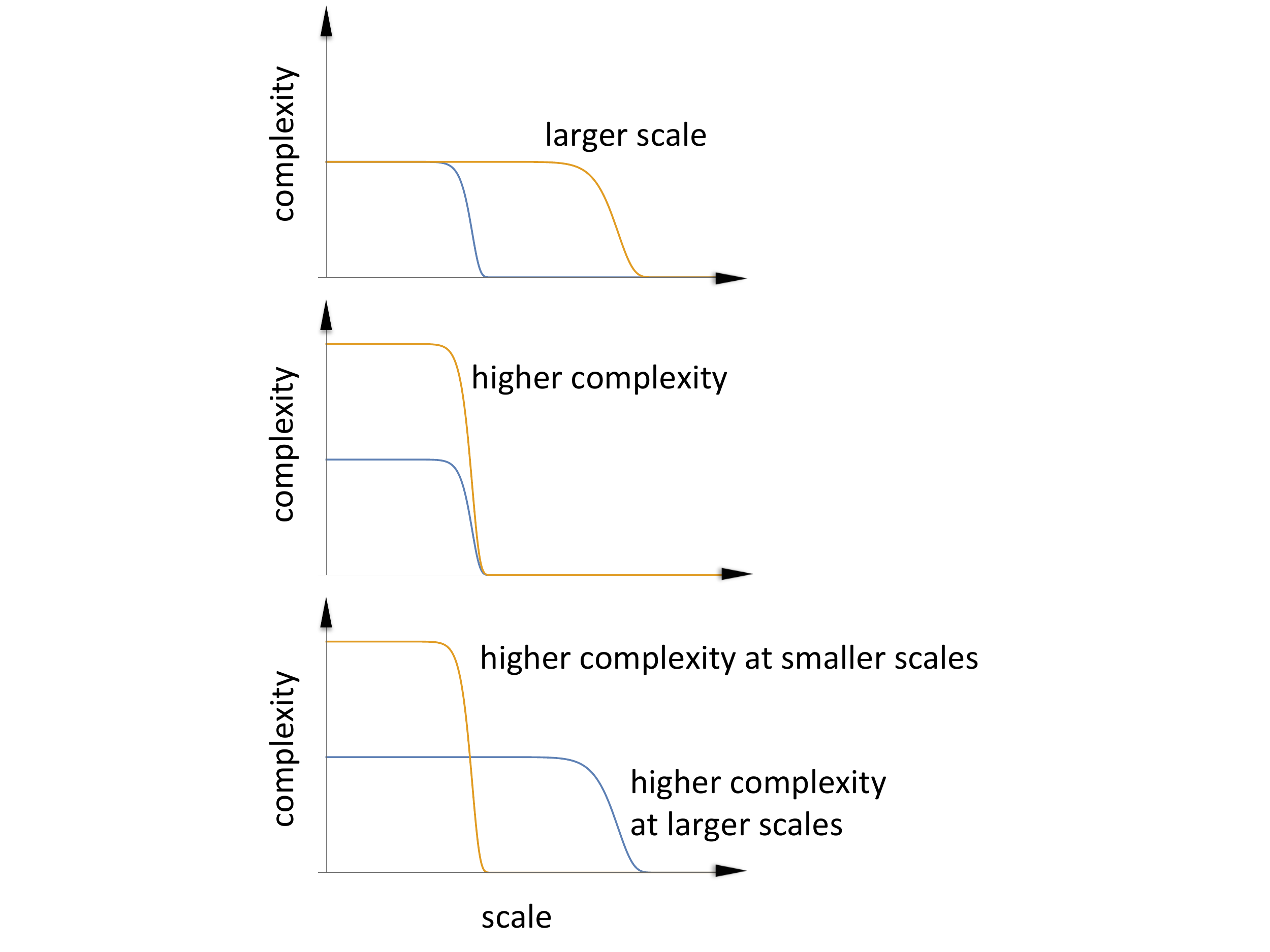}
\caption{Schematic complexity profiles of militaries in conflict.  Top: If two armies are operating with the same number of possible behaviors but at different scales, the larger-scale one is favored.  Middle: If two armies are operating at the same scale but with different numbers of possible behaviors, the higher-complexity one is favored.  Bottom: If two armies are operating at different scales and with different numbers of possible behaviors, which one is favored depends on the terrain (see text).  Note that these profiles are simplified to highlight the key concepts; actual militaries operate at multiple scales.  More generally, the top and middle graphs depict conflicts in which one army has at least as much complexity as the other at every scale.}
\label{fig:military}
\end{center}
\end{figure}

Since complexity is defined only with respect to a particular scale, we can refine the Law of Requisite Variety: To be effective, a system must match (or exceed) the complexity of the environmental behaviors to which it must differentially react at all scales for which these behaviors occur~\cite{allen2017multiscale}.  To illustrate this multi-scale version of the Law of Requisite Variety, we consider military conflict~\cite{military} (see \cref{fig:military}).  Here, one military can be considered as the system, while the other military is part of the environment with which the system must interact.  For two militaries of equal complexity, i.e. with the same number of behaviors, but with one military operating at a larger scale (e.g. two very tightly controlled armies, but with one army larger than the other), the larger-scale military will likely win.  For two militaries of equal scale but unequal complexity (e.g. two equally sized and equally powered fleets, but with one being more maneuverable than the other), the higher-complexity military will likely win, since the high-complexity military has an action for every action of the lower-complexity military but not vice versa.  When a military with high complexity at a smaller scale (e.g. a guerrilla force) conflicts with a military with larger-scale behavior but lower complexity (e.g. the U.S. army in Vietnam or the Soviet army in Afghanistan), the terrain, which constrains the scale of the conflict, plays an important role.  In an open field, or in open waters, the military that has more complexity at the larger scales is favored, while in the jungle or in the mountains, higher complexity at smaller scales is favored.    

As another example, healthcare involves both small-scale tasks with high overall complexity such as case management, as well as large-scale, lower complexity tasks, such as manufacturing and delivering vaccines~\cite{baryamhealth}.  (Delivering vaccines is lower complexity but higher scale because the same actions will be performed for nearly everyone.)  Large-scale top-down organizations and initiatives are suited for large-scale, lower complexity tasks, but tasks like case management require health systems with a high degree of small-scale (i.e. local) complexity.  

The eurozone provides a potential illustration of a multi-scale complexity mismatch.  Fiscal policy is made predominantly at the scale of individual countries and thus has a higher complexity at the country scale but relatively little complexity at the scale of the entire eurozone, while monetary policy is made at the scale of the entire eurozone and thus has some complexity at the scale of the eurozone but lacks the ability to vary (i.e. lacks complexity) at the scale of individual countries.   Many have argued that economic difficulties within the eurozone have arisen because this mismatch has precluded effective interactions between fiscal and monetary policy~\cite{canzoneri2005monetary,semmler2004monetary,alessandrini2015absence,dan2014euro,drudi2012interplay}.

Problems arise not from too much or too little complexity (at any scale) per se but rather from mismatches between the complexities of a task to be performed and the complexities of the system performing that task.\footnote{Incidentally, human emotions appear to reflect this principle: we are bored when our environment is too simple and overwhelmed when it is too complex~\cite{bored}.}
Note that the system in one scenario may be the task/environment in another; for instance, the same complexity that helps a system interact with its environment may prevent its effective management by other systems.
In none of the above examples have the complexity profiles been precisely calculated, nor have scales been precisely defined.  Instead, proxies for scale are used and estimated comparisons of complexity made.  Such an approach cannot yield precise results (indeed, no approach can, given the complexity a full description of such systems would require), but additional precision is not needed when even the approximate analysis reveals large mismatches in complexity.\footnote
{Just as this analysis of the space of possible behaviors 
can be used even in the face of uncertainty regarding a system's precise mechanisms and outcomes, physicists can use the property of entropy (sometimes considering how quantities related to entropy change across scale) to classify phase transitions even when they cannot, from first principles, determine precise quantities such as the amount of heat generated by a phase transition or the temperature at which it occurs (such quantities must be determined empirically).}
(To remedy the diagnosed mismatches, more detailed analyses may be required.)   While it may be tempting to attribute the problems arising from a complexity mismatch to particular proximate causes and chains of events, problems of one form or another will be inevitable unless the underlying mismatch is addressed.

\subsection{Subdivided systems}
\label{sec:subdivided}
Even if the complexity of the system matches that of its environment at the appropriate scales, there is still the possibility of a complexity mismatch.  Consider two pairs of friends---four people total, each of whom can lift 100 pounds---and consider two 200-pound couches that need to be moved.  Furthermore, assume that each person is able to coordinate with her friend but not with either of the other two people.  Overall then, the system of people has sufficient complexity at the appropriate scales to move both couches since each pair of friends can lift one of the 200-pound couches.  However, were one person from each pair of friends to be assigned to each couch, they would not be able to lift the couches because the two people lifting each couch would not belong to the same pair of friends and thus would not be able to coordinate their actions.  The problem here is that while the pairs of friends possess enough overall complexity at the right scales to lift the couches, the subdivision within the system of friends is not matched to the natural subdivision within the system of couches.  The mismatch in complexity can be seen if we focus our attention on just a single couch: while the couch requires coordinated action at the scale of 200 pounds, the two people lifting it are capable only of two independent actions, each at the scale of 100 pounds.

The way in which academic departments are organized provides a more realistic example of the potential of subdivision mismatch.  Academia has multiple levels of subdivision (departments, subfields, etc.) in order to organize knowledge and coordinate people, resulting in a high overall degree of complexity across multiple scales, where scale could refer to either the number of coordinated people or the amount of coordinated knowledge, depending on which aspect of the academic system is under consideration.  Similarly, there are multiple levels of natural subdivision in the set of problems that academia can potentially address, with each subdivision of problems requiring particular types of coordinated knowledge and effort in order to be solved.  Academia's complexity across multiple scales allows it to effectively work on many of these problems.  However, there may exist problems that academia, despite having sufficient overall multi-scale complexity, is nonetheless unable to solve because the subdivisions within the problem do not match the subdivisions within academia.  The increase in interdisciplinary centers and initiatives over the past few decades suggests the perception of such a mismatch; however, the structure of the academic system as a whole may still hinder progress on problems that do not fall neatly within a discipline or sub-discipline~\cite{nissani1997ten,brewer1999challenges,feller2002new,Rhoten2004,lele2005practicing,jacobs2009interdisciplinarity}.

The above examples provide an illustration of the principle that in order for a system to differentially react to a certain set of behaviors in its environment, not only must the system as a whole have at least as much complexity at all scales as this set of  environmental behaviors (as described in \cref{sec:ashby}), but also \textit{each subset} of the system must have at least as much complexity at all scales as the environmental behaviors corresponding to that subset.  A good rule of thumb for applying this principle is that decisions concerning independent parts or aspects of a system
should be able to be made independently, while decisions concerning dependent parts of the system should be made dependently.  It follows that the organizations that make such decisions should be subdivided accordingly, so that their subdivisions match the natural divisions in the systems with which they interact.\footnote
{The subdivisions present in the human brain and the analysis of subdivisions in neural networks more generally~\cite[Chapters~2.4-2.5]{textbook} demonstrate how systems that are subdivided so as to match the natural subdivisions in their environments outperform those with more internal connectivity.}

\subsection{Hierarchies}
\label{sec:hierarchies}
\begin{figure}
\begin{center}
\includegraphics[width=.5\textwidth]{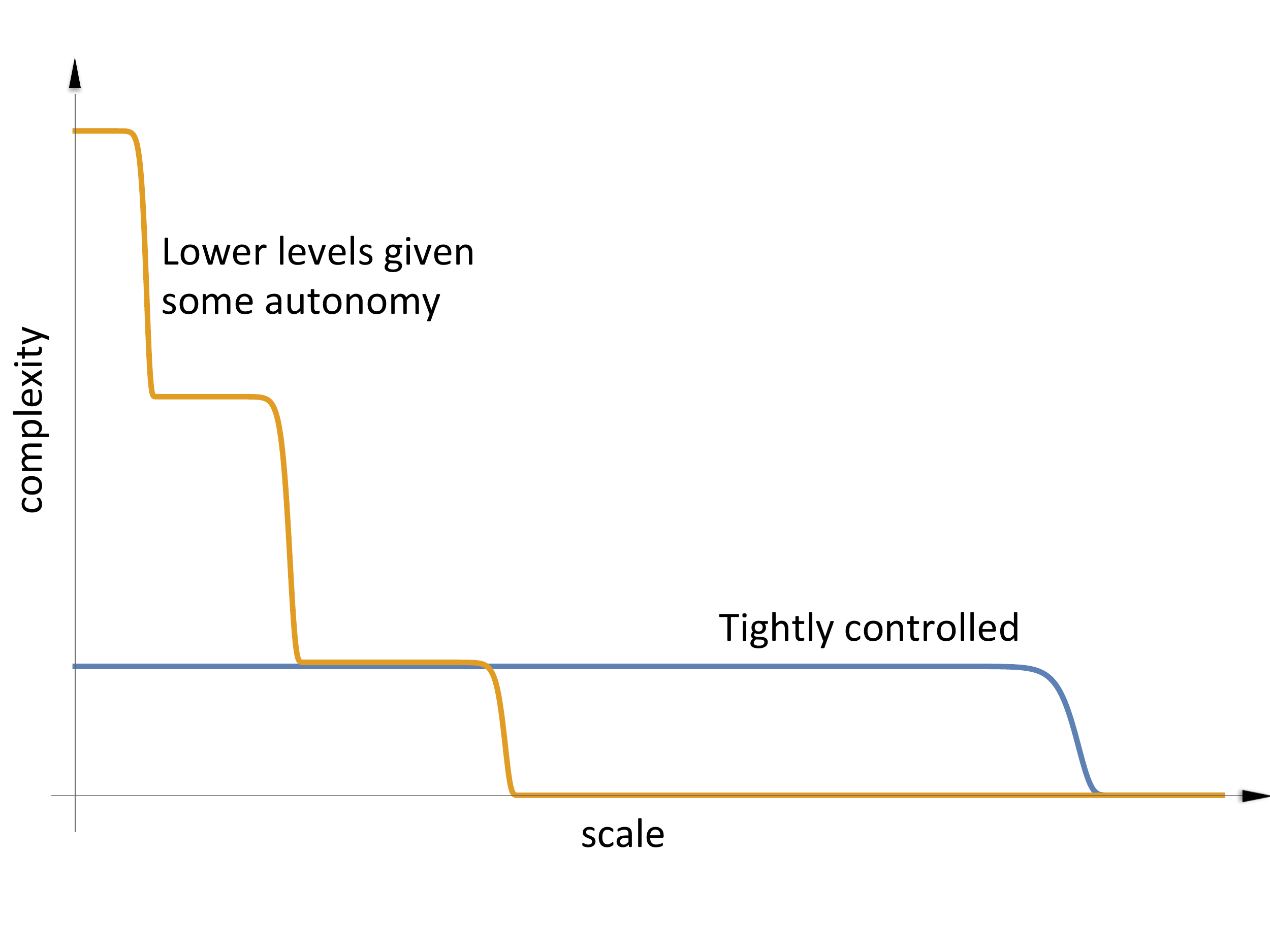}
\caption{Complexity profiles of two hierarchies, each with the same number of people.  Here, the scale is the number of coordinated man-hours.  In one hierarchy, all decisions, regardless of the scale, are made by a single person, while in the other, different decisions are made at various levels of the hierarchy.}
\label{fig:hierarchies}
\end{center}
\end{figure}

A common way in which systems are organized is through hierarchies.  In an idealized hierarchy, there are no lateral connections: any decision that involves multiple components of the hierarchy 
must pass through a common node under whose control these components all (directly or indirectly) lie.  The complexity profile of such a hierarchy depends on the rigidity of the control structure (\cref{fig:hierarchies}).  At one extreme, every decision, no matter how large or small, is made by those at the top of the hierarchy.
This hierarchy has the same amount of complexity across all its scales: namely the complexity of whatever decisions are being made at the top.  At the other extreme, there is no communication within the hierarchy, and every individual acts independently.  This hierarchy has very little complexity beyond the individual level.  Between these two extremes is a typical hierarchy, in which 
different decisions are made at different levels.  

No type of hierarchy is inherently better than any other. For a particular environment, the best hierarchy is one for which the complexity profile matches that of the tasks needed to be performed.  A tightly controlled (top-heavy) hierarchy is not well suited to environments in which there is a lot of variation in the systems with which the lower levels of the hierarchy must interact; neither is a very loosely controlled hierarchy well suited to environments that require large-scale coordinated action.  For example, centralizing too much power within the U.S. governance system at the federal (as opposed to the local or state) level would not allow for sufficient smaller-scale complexity to match the variation among locales; too decentralized a governance system would not allow for sufficient larger-scale complexity to engage with problems that require nationally coordinated responses.\footnote
{We can consider not just the overall complexity profile of governance systems but also how well the subdivisions in governance systems match those within their territories (\cref{sec:subdivided}).   Metropolitan areas are in some ways more similar to one another than they are to the rural areas of their respective states.  So while dividing the U.S. into 50 states provides substantial lower-scale governmental complexity, this complexity is not necessarily well matched to natural urban-rural divides.  To the extent that such a mismatch exists, there may be issues currently handled at the state level that would be better handled at the local level, thereby allowing for different policies in urban and rural areas (and likewise, perhaps some of the powers that some argue should be devolved from the federal to the state level should in fact be devolved to the local level).}  
Assigning decisions to higher levels in hierarchies allows for more efficiency and scale, but such decisions result in less adaptability because when they are incorrect, they affect more of the system and---as larger-scale changes tend to require longer time-scales to enact---are more difficult to roll back.

It is important to distinguish between the complexity of a hierarchy and the complexity of the decisions that the people within the hierarchy are capable of making.  For instance, one could design a tightly controlled hierarchy that could take a large number of large-scale actions (i.e. high complexity at its largest scale), but since the decision-making abilities of even the most capable humans are of finite complexity, 
the individuals at the top may be fundamentally unable to correctly choose from among these actions.   This brings us to an important limitation of hierarchies: the complexity of the decisions concerning the largest-scale behaviors of a hierarchy---the behaviors involving the entire organization---is limited by the complexity of the group of people at the top~\cite{behaviors}.  
Thus, a hierarchy will necessarily fail when the complexity of matching its largest-scale behaviors to those of its environment\footnote
{Note that the complexity of deciding which behaviors of a system should correspond to which behaviors of its environment is generally much greater than the complexity of either the system or the environment alone: for example, if both the system and environment have 10 possible behaviors, the system has enough complexity to match the environment, but properly deciding which behaviors of the system should correspond to which environmental conditions requires correctly choosing one option out of a space of 3,628,800 (10 factorial) possibilities.  The space of possible behaviors of a system and its environment may be much smaller than the space of possible decisions concerning the management of the system's actions in its environment.}
is higher than the complexity of decision-making that is achievable by any individual or committee.  The failure of command economies provides a stark example: the allocation of resources and labor is too complex a problem for any one person or group of people to understand.  Markets allocate resources via a more networked system: decisions regarding how to allocate resources are made without any individual making them, just as decisions are made in the human brain without any neuron making them.  (Whether or not these market allocations are desirable depends in part on the way in which the market is structured and regulated.)

We began by considering idealized hierarchies with only vertical connections, but lateral connections provide another mechanism for enabling larger-scale behaviors.  For instance, cities can interact with one another (rather than interacting only with their state and national governments) in order to copy good policies and learn from each other's mistakes. Through these sorts of evolutionary processes (described further in \cref{sec:decision}), large-scale decisions (large-scale because policies may be copied by multiple cities) that are more complex than any individual component can be made.  Such lateral connections can exist within a hierarchical framework in which the top of the hierarchy (in this example, the national government) maintains significant control, or they can exist outside of a hierarchical structure, as in the human brain.  Furthermore, these lateral connections can vary in strength. Overly strong connections lead to herd-like behaviors with insufficient smaller-scale variation, such as groupthink~\cite{janis2008groupthink,trueman1994analyst,pan2012decoding} (no system is exempt from the tradeoff described in \cref{sec:tradeoffs}), while overly weak connections result in mostly independent behavior with little coordination.

\section{Analyzing Complex Systems}
\label{sec:analyze}
The previous section has examined some of the general properties of systems with many components.  But how do we study particular systems?  How do we analyze data from complex systems, and how do we choose which data to analyze?   

\begin{figure}
\begin{center}
\includegraphics[width=.5\textwidth]{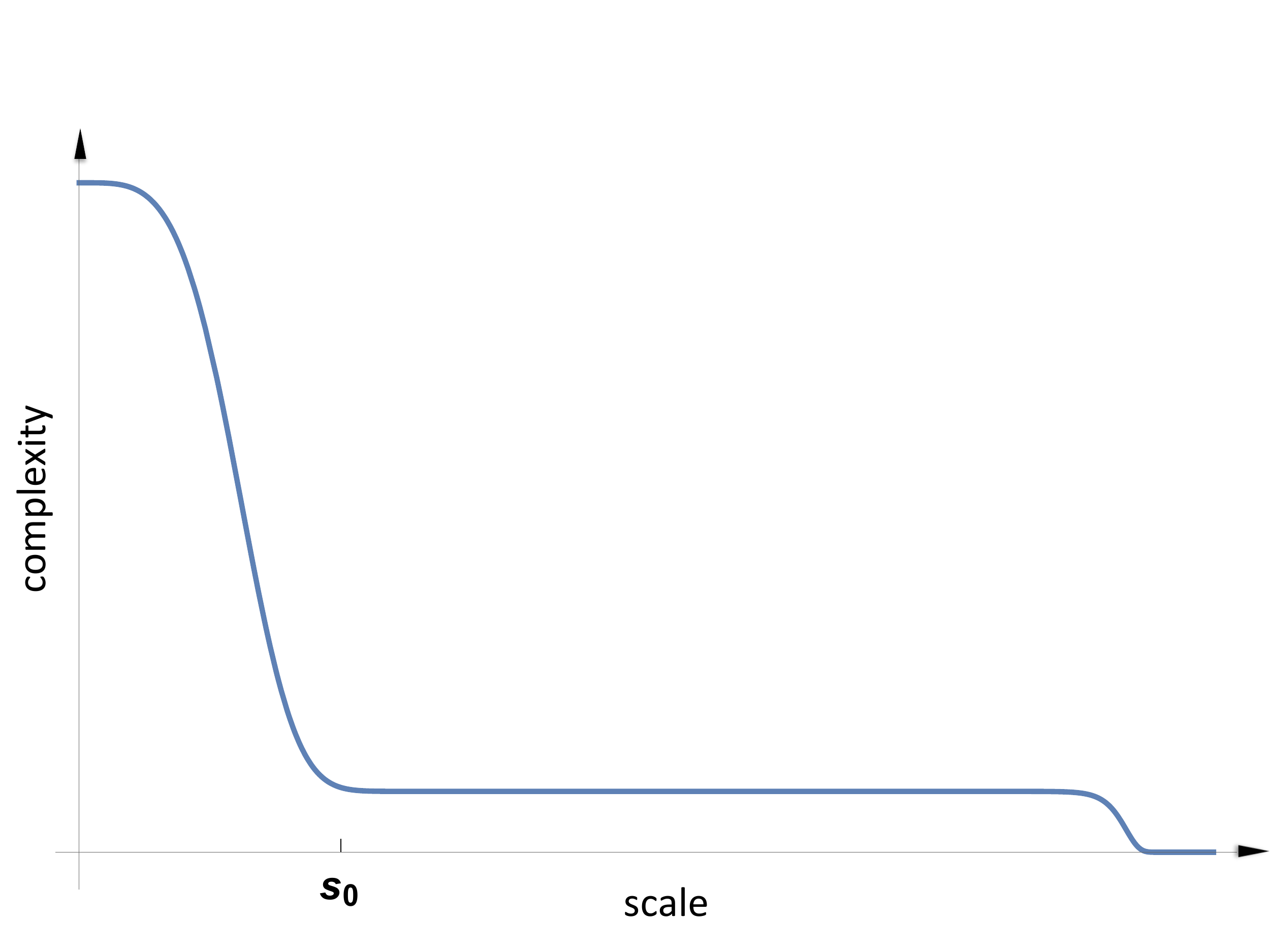}
\caption{A complexity profile of a system with a separation of scales. A separation of scales implies that the behaviors occurring below a certain scale ($s_0$ in the above figure) are at larger scales mostly independent from one another and that therefore, at these larger scales, only the average effects of the small-scale behaviors are relevant.}
\label{fig:sep}
\end{center}
\end{figure}

\subsection{How do we understand any system?}
\label{sec:mft}
In a sense, it is surprising that we can understand any macroscopic system at all, as even a very simple mechanical system has trillions upon trillions of molecules.  We are able to understand such systems because they possess a \textit{separation of scales}~\cite{bigdata}, meaning 
that the macroscopic behavior we are interested in occurs at a far larger scale than the behavior of the individual molecules, with not much behavior occurring in between these two scales (see \cref{fig:sep}).  This separation allows us to treat the macroscopic and microscopic behaviors separately: for mechanical systems, we treat the macroscopic behavior explicitly with Newtonian mechanics, while the microscopic behavior is considered in aggregate using thermodynamics. 

More generally, the approach described above is an example of a \textit{mean-field theory}~\cite{Kadanoff2009}, in which the average behaviors of a system's components are explicitly modeled and the deviations of the individual components from this average are treated as statistically independent random fluctuations.  This approach works very well for systems such as computers, cars, airplanes, and buildings, 
in which the motions of individual molecules are---apart from some mostly uncorrelated fluctuations---well described by the motion of the piece of material to which they belong. 
Mean-field assumptions are also often employed in analyses of biological, social, and economic systems; these assumptions work well in many cases, but, as we will see, they are not always appropriate for complex systems.  It is important, therefore, to determine under what conditions mean-field theory holds.

\subsection{When mean-field theory breaks down}
\label{sec:fail}
The systems for which mean-field theory applies exhibit large-scale behaviors that are the average of the behaviors of their components.  They must possess a separation of scales, which arises when the statistical fluctuations of their components are sufficiently independent from one another above a certain scale.
Mean-field theory may hold even in the presence of strong interactions, so long as the effect of those strong interactions can be captured by the average behavior of the system---that is, so long as each component of the system can be modeled as if it were interacting with the average (i.e. mean field) of the system.  For example, the large-scale motion of solids is well described by mean-field theory, even though the molecules in a solid interact with one another quite strongly, because the main effect of these interactions is to keep each molecule at a certain distance and orientation from the average location (center of mass) of the solid.  Likewise, under some (but certainly not all) conditions, economic markets can be effectively described by modeling each market actor as interacting with the aggregate forces of supply and demand rather than with other individual market actors.  

However, when there are sufficiently strong correlations between the components of the system, i.e. when the interactions between a component of the system and a specific set of other components (as opposed to its general interaction with the rest of the system) cannot be neglected, mean-field theory will break down.\footnote
{Incidentally, the failure of mean-field theory to describe certain physical phase transitions led physicists to develop a new, multi-scale approach (the renormalization group), a foundation of much of complex systems science.}  
These systems will instead exhibit large-scale behaviors that arise not solely from the properties of individual components but also from the relationships between components.  For example, while the behavior of a muscle can be roughly understood from the behavior of an individual muscle cell, the behavior of the human brain is fundamentally different from that of individual neurons, because cognitive behaviors are determined largely by variations in the synapses \textit{between} neurons.  Similarly, the complex ecological behaviors of a forest cannot be determined by the behaviors of its constituent organisms in isolation.

Because their small-scale random occurrences are not statistically independent, complex systems often exhibit large-scale fluctuations not predicted by mean-field theory, such as forest fires, viral content on social media, and crashes in economic markets. 
Sometimes, these large-scale fluctuations are adaptive: they enable a system to collectively respond to small inputs~\cite{starlings}.  (For instance, humans respond strongly to minor disturbances in the density of air, such as the sound of their own names.)  However, these large-scale fluctuations sometimes pose systemic risks.

\begin{figure}
\begin{center}
\includegraphics[width=.5\textwidth]{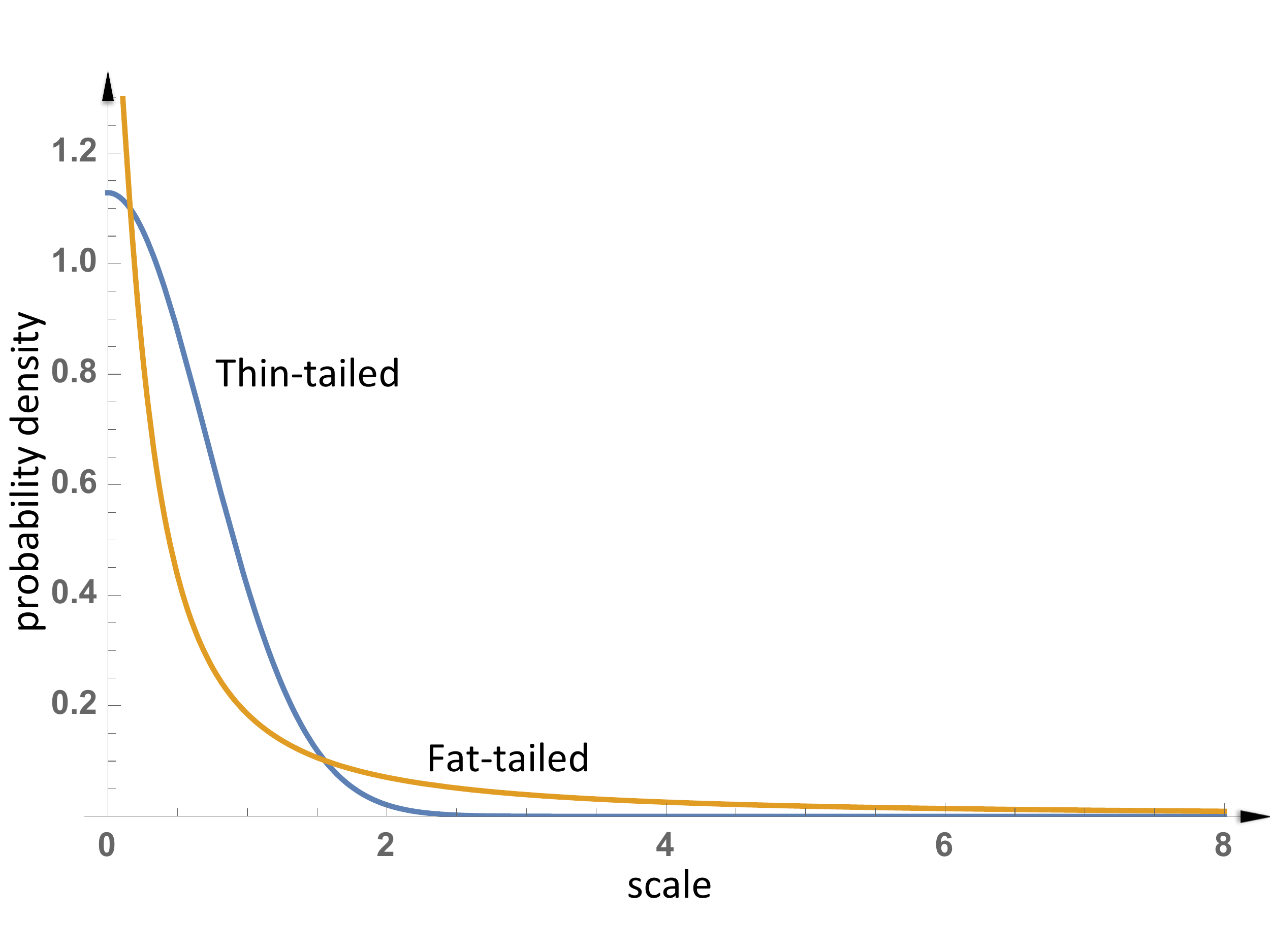}
\caption{A normal distribution (thin-tailed) and a distribution with a power-law decay (fat-tailed). The fat-tailed distribution may appear more stable, due to the lower probability of small-scale fluctuations and the fact that samples from the distribution may not contain any extreme events.   However, sooner or later, a fat-tailed distribution will produce an extreme event, while one could wait thousands of lifetimes of the universe before a normal distribution produces a similarly extreme event.  Note that the axes of this graph are truncated; the illustrated fat-tailed distribution can, with small but non-negligible probability (0.04\%), produce events with a scale of one million or more.}
\label{fig:tails}
\end{center}
\end{figure}

\subsection{Fat-tailed distributions and systemic risk}
\label{sec:risk}
When the components of a system are independent from one another above a certain scale, then at much larger scales, the magnitude of the fluctuations of the system follow a normal distribution (bell curve),\footnote
{This follows from the central limit theorem.}
for which the mean and standard deviation are well-defined and for which events many standard deviations above the mean are astronomically improbable.  Interdependencies, however, can lead to a distribution of fluctuations in which the probability of an extreme event, while still small, is not astronomically so.  Such distributions are characterized as \textit{fat-tailed}---see \cref{fig:tails}.  For example, while human height follows a thin-tailed distribution, with no record of anyone over twice as tall as the average human, human wealth---due to the complex economic interactions between individuals---follows a fat-tailed distribution, with multiple individuals deviating from the average by factors of more than one million~\cite{davies2017estimating}.

One danger of interdependencies is that they may make systems appear more stable in the short term by reducing the extent of small-scale fluctuations, while actually increasing the probability of catastrophic failure~\cite{buldyrev2010catastrophic,vespignani2010fragility,helbing2013globally,balsa2020deglobalization}.  This danger is compounded by the fact that when underlying probability distributions have fat tails (a situation made more likely by interdependencies), standard statistical methods often break down, leading to potentially severe underestimates of the probabilities of extreme events~\cite{Taleb2019}.\footnote{For instance, the mean and variance of a fat-tailed distribution may not be well-defined, and even if they are, they may not be able to be reliably estimated from a finite sample due to the likelihood of a single data point or lack thereof substantially skewing the estimate.}  As a thought experiment, imagine 100 ladders, each with a 1/10 probability of falling.  If the ladders are independent from one another, the probability that all of them fall is astronomically low (literally so: there is about a $10^{20}$ times higher chance of randomly selecting a particular atom out of all of the atoms in the known universe).  If we tie all the ladders together, we will have made them safer, in the sense that the probability of any individual ladder falling will be much smaller, but we will have also created a non-negligible chance that all of the ladders might fall down together.  Other examples include the interconnectedness of our financial systems resulting in the possibility of global market crashes~\cite{May2008,harmon2010networks,haldane2011systemic,farmer2012complex,schweitzer2009economic,sornette2017stock} and the interconnectedness of travel routes increasing the probability of pandemics such as the Spanish flu and COVID-19~\cite{rauch2006,siegenfeld2020eliminating}.  When such crises do occur, they are often attributed to proximate causes or chains of events, and measures are then implemented to ensure that those particular chains of events will not occur again.  But unless the underlying systemic instabilities are addressed, another crisis is bound to happen sooner or later, even if its precise form cannot be predicted.  
 
\subsection{Understanding complex systems} 
Because it is usually easier to collect data regarding components of a system than it is to collect data regarding interactions between components, studies often fail to capture the information relevant to complex systems, since complex large-scale behaviors critically depend on such interactions.  Furthermore, as discussed in \cref{sec:risk}, data analysis can severely underestimate the probability of extreme events (tail risk).  Finally, analyses often (implicitly) assume linearity, i.e. they assume that the total impact of a set of factors is equal to the sum of the impacts of each individual factor, an assumption that often breaks down for complex systems, which may possess feedback loops, abrupt transitions (tipping points), and other highly nonlinear behaviors~\cite{ludwig1997sustainability,heng2008conflict,scheffer2009early,scheffer2012anticipating,helbing2015saving,rutter2017need,wiesner2018stability,siegenfeld2020negative}.

How can we understand the systems for which these standard approaches do not apply?  Our understanding of \textit{all} systems with many components depends on \textit{universality}~\cite{Kardar2007}, i.e. the existence of large-scale behaviors that do not depend on the microscopic details.  The standard approaches are predicated on the assumption of sufficient independence between components, which allows large-scale behaviors to be determined without a full accounting of the system's details via mean-field theory.\footnote{Formally, the statistical fluctuations of the components must, above a certain scale, be sufficiently independent so as to satisfy the assumptions of the central limit theorem.  The central limit theorem is the manifestation of universality that explains the ubiquity of normal distributions.}  But mean-field theory is just one example of universality.  

Sound is another example: 
all materials, regardless of their composition, allow for the propagation of sound waves.   Sound behaves so similarly in all materials because at the length scales relevant to sound waves, which are far larger than the sizes of individual atoms and molecules, the effect of the microscopic parameters is merely to set the speed of the sound.\footnote
{For Quantum Electrodynamics (the theory of how light and electrons interact), we \textit{still do not know} the microscopic details.  Yet we can nonetheless make predictions accurate to ten decimal places because, as can be shown with renormalization group theory, the only effect of these microscopic details at the scales at which we can make measurements is to set the electron mass and charge---quantities that, like the speed of sound in any particular material, can be measured (but not predicted).}   
 Note that sound waves cannot be understood as a property of the average behavior---in this case, average density---of a material, since it is precisely the systematic correlations in the deviations from that average that give rise to sound.  Nor is sound best understood by focusing on the small-scale details of atomic motion: scientists understood sound even before they learned what atoms are. 
   The key to understanding sound waves is to recognize that they have a multi-scale structure---with larger-scale fluctuations corresponding to lower frequencies and smaller-scale fluctuations corresponding to higher frequencies---and to model them accordingly.

\begin{figure}
\begin{center}
\includegraphics[width=.5\textwidth]{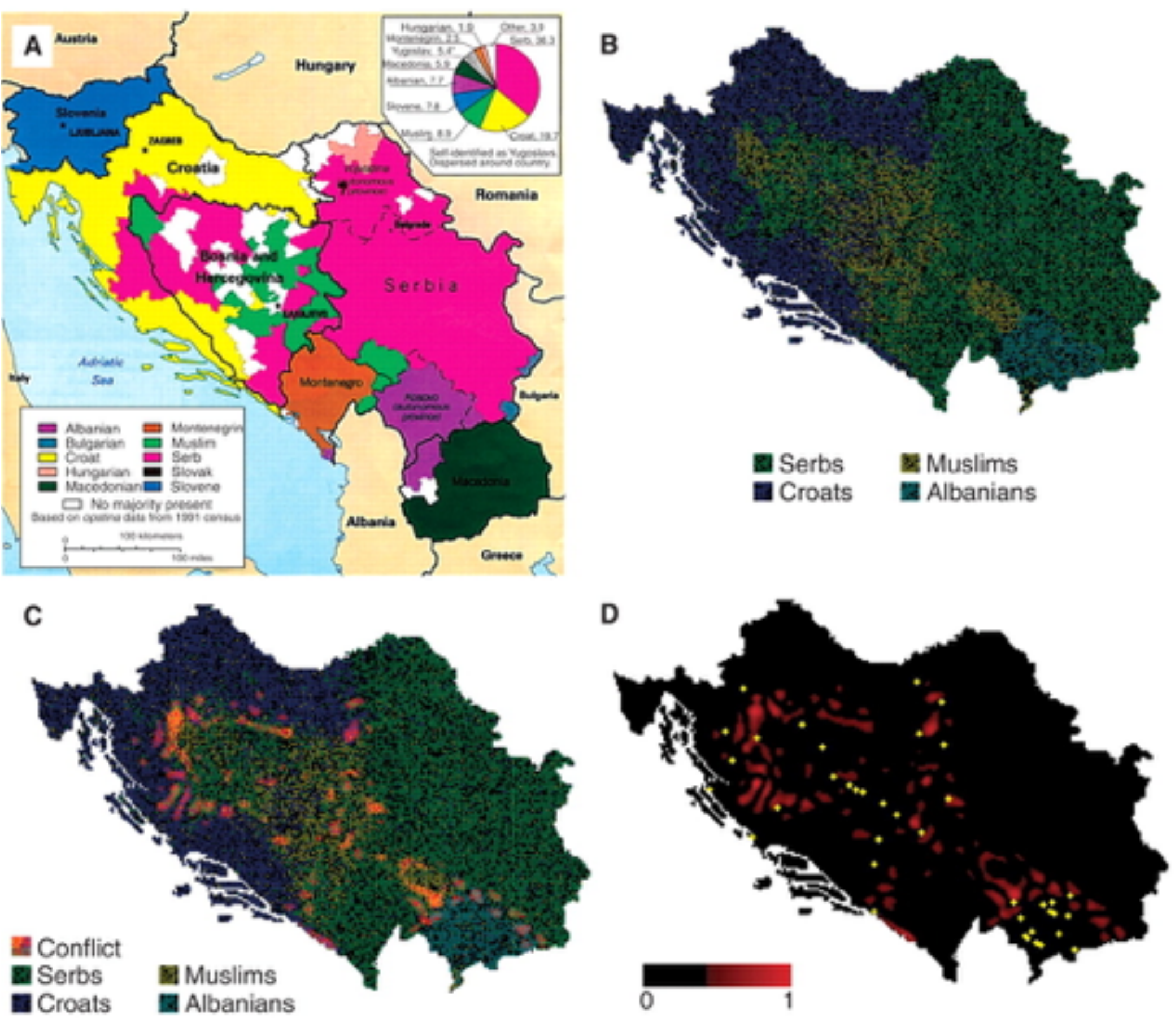}
\caption{A figure from Lim \etal's paper on ethnic violence~\cite{ethnic}.  The sites where their model predicts a potential for ethnic violence are shown in red in panels C and D, with confirmed reports of ethnic violence depicted by the yellow dots in panel D.}
\label{fig:ethnic}
\end{center}
\end{figure}

Lim \etal~apply this approach to studying ethnic violence~\cite{ethnic}.  They built a predictive model to analyze where ethnic violence has the potential to occur and applied their model to India and to what was Yugoslavia.  Ethnic violence has many causes, but rather than focusing on specific, culturally dependent mechanisms or on the average properties of regions, such as demographic or economic statistics, the authors instead considered the multi-scale patterns in how 
ethnic groups were geographically distributed (\cref{fig:ethnic}).  They found that ethnic violence did not occur when the ethnic groups were either well mixed or well separated but rather occurred only when ethnic groups separated into geographic patches,\footnote
{This separation falls into the same universality class as the separation of oil and water.}
with the violence most likely to occur for geographic patches of a particular size.\footnote
{This analysis implies that ethnic violence can be prevented by the use of well-placed political boundaries, as in Switzerland~\cite{rutherford2014}.}  
Although not explicitly included in the analysis, specific details of a region are relevant insofar as they are either a cause or an effect (or both) of the patch size.\footnote
{For instance, animosity between two ethnic groups, though not explicitly considered, may be a cause as well as a consequence of the geographic segregation~\cite{schelling1971}.}  

\begin{figure}
\begin{center}
\includegraphics[width=.5\textwidth]{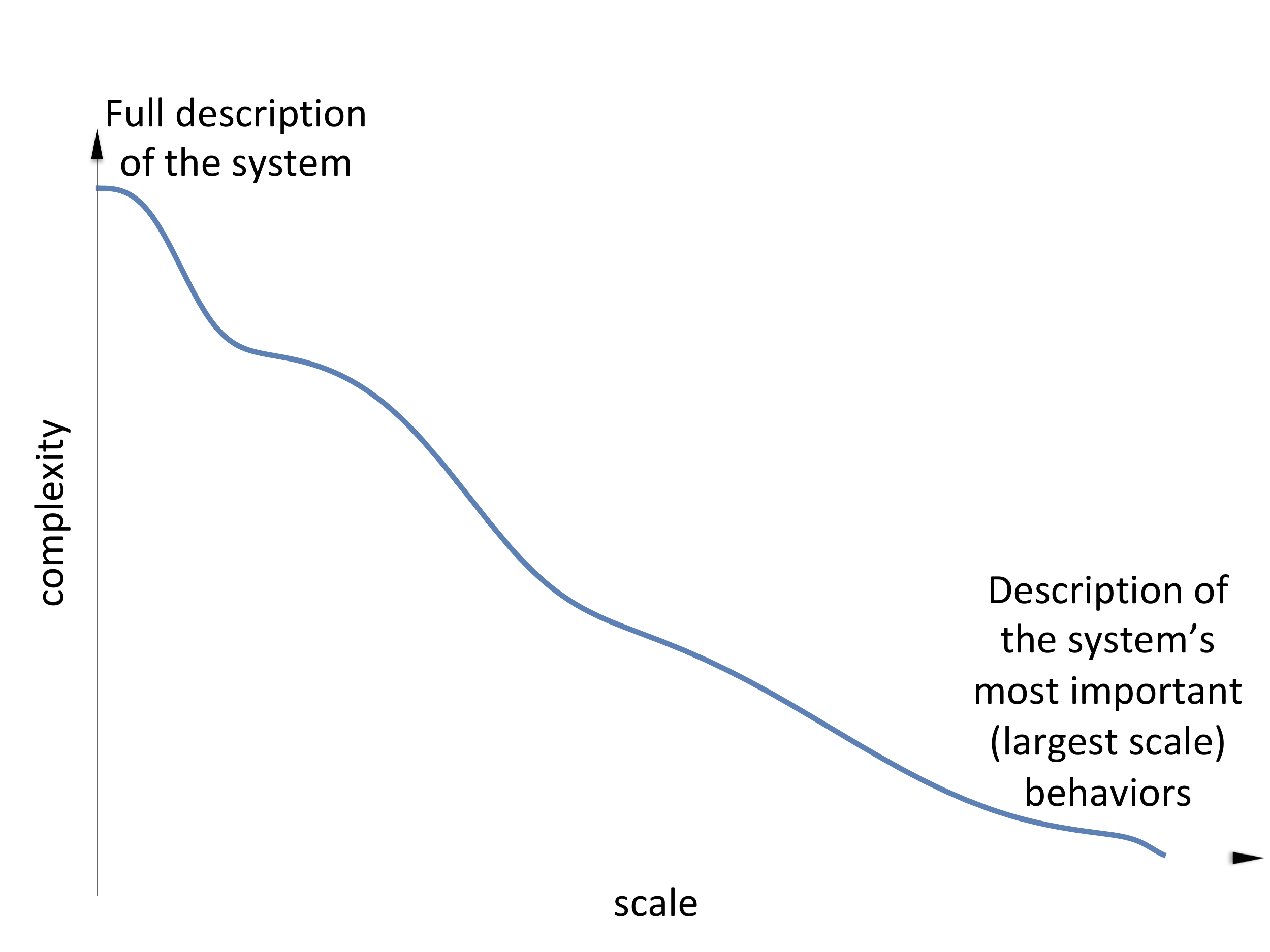}
\caption{A representative complexity profile of a complex system.  Understanding all the details (i.e. all of the small-scale behaviors) is impossible and unnecessary; the most important information is contained in the large-scale behaviors.  However, for systems for which mean-field theory does not apply, characterizing these behaviors will involve more than a simple average.}
\label{fig:infoscale}
\end{center}
\end{figure}

Understanding all of the details of any complex system is impossible, just as it is for most systems with a separation of scales; there is just too much complexity at the smallest scale.  But unlike the behaviors of systems with a separation of scales, the important large-scale behaviors of complex systems are not simply the average of their small-scale behaviors.  The interdependencies at multiple scales can make it difficult or impossible to precisely understand how small-scale behaviors give rise to larger-scale ones, but even for complex systems, there is much less complexity at the larger scales than there is at the smaller scales.  Thus, there will always be large-scale behaviors that do not depend on most of the system's details (see \cref{fig:infoscale}).  The key to analyzing these behaviors is to find the appropriate mathematical (or conceptual) description---i.e. to identify variables that describe the relevant space of possible (large-scale) behaviors---which for complex systems is not a simple average nor a full account of all the details.  For additional examples of this multi-scale approach, see ref.~\cite{bigdata}.

\section{Complex Systems and Uncertainty}
\label{sec:decision}
Although the principles discussed throughout \cref{sec:basic,sec:analyze} help us recognize the fundamental properties and limitations of systems, our understanding of most complex systems will inevitably be imperfect.  
And regardless of how well-considered a plan is, a truly complex system will present elements that were not considered ahead of time.\footnote
{It should also be noted that in a functional system with a high degree of complexity, the potential positive impact of a change is generally much smaller than its potential negative impact.  For example, a change to the wiring in a computer is unlikely to dramatically improve the computer's performance, but it could cause the computer to crash.  Airplanes are another example.  This phenomenon is a consequence of the fact that, by definition, a high degree of complexity implies that there are many system configurations that will not work for every one configuration that will.} 
Given the absence of perfect knowledge, how can the success of systems we design or are part of be assured?
While the success of many systems rests on the assumption that good decisions will be made, some systems do not depend on individual understanding and can perform well in spite of the fallibility of decision-makers (whether due to corruption, subconscious bias, or the fundamental limitations of human minds).  The study of complex systems approaches this observation scientifically by (implicitly or explicitly) considering the decision-makers themselves as part of the system and of limited complexity/decision-making ability. 
The question thus becomes: how do we design systems 
that exceed the complexity of the decision-makers within them?

\subsection{Evolutionary processes}
While uncertainty makes most systems weaker, some systems benefit from uncertainty and variability~\cite{turing2009computing,mcdonnell2011benefits,goldenfeld2011,antifragile}.  The common characteristic of these systems is their embodiment of some sort of evolutionary process, i.e. a process in which successful changes are copied (and further modified) while unsuccessful changes are not.  The classic evolutionary processes are biological: due to variability introduced by random mutations, organisms with the complexity and scale of humans evolved from single-celled organisms.  Furthermore, humans themselves have the property of benefiting from exposure to random shocks (provided the shocks are not too strong).   Immune system performance is improved by early exposure to non-lethal pathogens~\cite{olszak2012microbial,su2013virus}; muscles and bones are strengthened by micro-tears and micro-fractures, respectively; we learn by exposure to new information and problem-solving; and our psychologies are strengthened by exposure to adversity, provided the adversity is not too severe~\cite{seery2011resilience,lukianoff2015coddling}.

Competitive market economies provide another example of how systems can thrive on uncertainty.   Due to our ignorance of which will succeed, many potential innovations and businesses must be created and improved upon in parallel, the successful ones expanding and the unsuccessful ones failing.   The successful among these can then be improved upon in the same manner---with many approaches being applied at once---and so on.  (However, without effectively regulated multi-scale cooperative frameworks---see \cref{sec:multievo}---large-scale parts of the economic system may optimize for the wrong goals, settling into harmful societal equilibria~\cite{ostrom2010beyond,luna2020corruption}.)  

Likewise, the internal processes of large organizations may follow an evolutionary pattern in which small parts of the organization can fail and thus be improved upon; without such flexibility, the entire organization may fail at once in the face of a changing internal or external environment.  
In some cases the failure of the entire organization makes room for more effective organizations to take its place (assuming the economy is sufficiently decentralized and competitive so that the organization in question is not ``too big to fail'').   
The collapse of government is generally not one of those cases, however~\cite{gard2015}, so it is especially important that governance systems possess the flexibility to internally benefit from randomness and uncertainty.  Perhaps counterintuitively, not allowing small failures to occur may weaken systems in the long run by halting evolutionary processes and by creating interdependencies that lead to systemic risk (\cref{sec:risk}).

In order to thrive in uncertainty and exceed the complexity of individual decision-making, systems can incorporate evolutionary processes so that they, even if very limited at first, will naturally improve over time.  
The first step is to allow for enough variation in the system, so that the system can explore the space of possibilities.  Since a large amount of variation means a lot of complexity and complexity trades off with scale (\cref{sec:tradeoffs}), such variation must occur at smaller scales (in both space and time).  For example, in the case of governance, enabling each city to experiment independently allows for many plans to be tried out in parallel and to be iterated upon.  The opposite strategy would be to enact one national plan, the effects of which will not be able to be comparatively evaluated.

The second step is to allow for a means of communication between various parts of the system so that successful choices are adopted elsewhere and built upon (e.g. cities copying the successful practices of other cities).  Plans will always have unintended consequences; the key is to allow unintended consequences to work for rather than against the system as a whole.   The desire for direct control must often be relinquished in order to allow complexity to autonomously increase over time.\footnote{Systems can explicitly design only systems of lesser complexity since an explicit design is itself a behavior of the first system.  However, systems that evolve over time can become more complex than their designers.} 

\begin{figure}
\includegraphics[width=.5\textwidth]{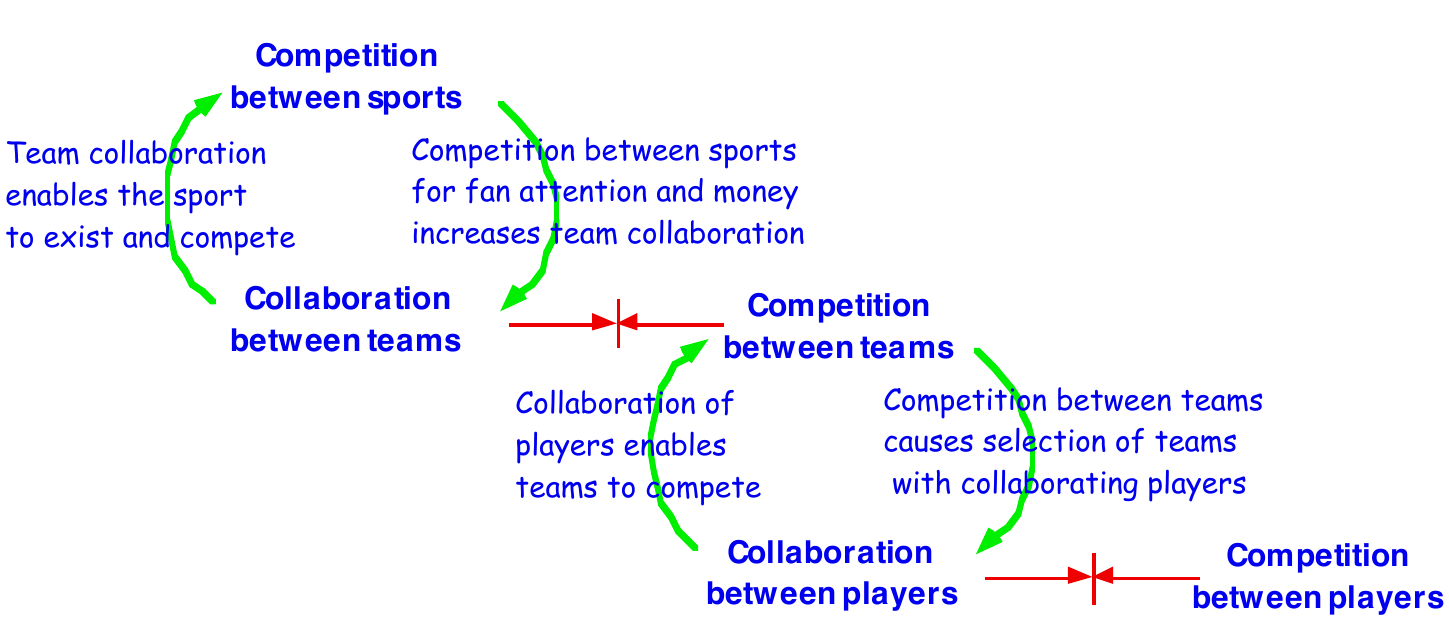}
\caption{An illustration from ref.~\cite[Chapter~7]{mtw} showing the interplay between cooperation and competition in the context of sports teams and leagues.}
\label{fig:sports}
\end{figure} 

\subsection{Multi-scale evolutionary processes}
\label{sec:multievo}
Successful evolutionary processes generally do not consist of unbridled competition but rather contain both competition and cooperation, each occurring at multiple scales~\cite{bar2006engineering}.  For example, cells cooperate within multicellular organisms in order to more effectively compete with other organisms, and organisms cooperate both within and between species in order to more effectively compete against other species.  Competition at larger scales naturally breeds cooperation at smaller scales because in order for a group to effectively compete against another group (large-scale competition), there must be cooperation within the group.  
Cooperation can also breed competition since sometimes the best way for the group to achieve its shared goals is to facilitate some healthy competition among its subgroups.    
Those subgroups must foster cooperation within themselves in order to effectively compete with each other, and they too may be able to increase the effectiveness of their internal cooperation by introducing some healthy competition among their members (\cref{fig:sports} provides an example).  If these members are themselves groups, the process of competition begetting cooperation that begets more competition can continue to even smaller scales.  This process can work in reverse as well: in order for individuals to compete more effectively, they may cooperate with each other to form groups, which in turn may cooperate to form even larger groups, and so on.  Thus, a complex network of cooperation and competition among groups of various sizes (scales) can naturally evolve.

In order to promote effective group cooperation, competition must be properly structured.  A soccer team in which the players compete with their own team members to score goals will not be effective, but one in which the players compete for the title of the most fit may be.  The framework in which competition occurs must be structured so that the competitors are incentivized to take actions that are net good for the group;
otherwise a kind of tragedy-of-the-commons situation occurs.  The potential for competition to go awry highlights the importance of having a multi-scale structure with competition occurring on multiple levels, rather than having everyone in the system compete with everyone else.  With the multi-scale structure, groups with unhealthy evolutionary dynamics are selected against, while groups with a healthy mix of competition and cooperation that benefits the entire group are selected for.\footnote
{There is evidence that the geographic nature of evolution---in which organisms evolve in somewhat separated environments and mean-field theory does not apply---has resulted in precisely this multi-scale structure and has therefore allowed for the evolution of genuine (e.g. not reciprocal) altruistic behavior~\cite{evo,wilson2015does}.}  
Market economic systems are successful not because free markets produce optimal outcomes (real-world markets often sharply deviate from the assumptions of free-market models, and externalities abound) but rather because, at their best, appropriately regulated market systems allow for multi-scale evolutionary processes to naturally arise, resulting in innovations and complexity far beyond what anyone could have imagined, let alone designed.

\section{Further reading}
\label{sec:further}
Complex systems science, also known as complexity science, contains many subfields.  
One starting point for exploring complex systems more broadly is this clickable map~\cite{map} of complex systems science and related fields.   Encyclopedias~\cite{lifesupportsystems,meyers2009encyclopedia} and textbooks~\cite{simon2019sciences,textbook,mitchell2009complexity,miller2009complex,sayama2015introduction,thurner2018introduction} provide a range of perspectives.  In addition to the topics and references discussed throughout this introduction, we provide a selection among the many works applying complex systems science to social systems and policy~\cite{byrne1998complexity,lempert2002new,bankes2002tools,glouberman2002complicated,teisman2008complexity,maroulis2010complex,hall2010complex,bai2010urban,cairney2012complexity,ball2012society,lipsitz2012understanding,geyer2015handbook,bar2015complexity,edmonds2015modelling,stroh2015systems} and management~\cite{weick1995sensemaking,stacey1996complexity,hock1999birth,hammer2009reengineering}.  Complex systems science includes, among others, the fields of system dynamics~\cite{sterman2010business}, evolutionary dynamics~\cite{kauffman1993origins,sober1999unto,axelrod2006evolution}, network science~\cite{barabasi2016network}, fractals and scaling~\cite{de1979scaling,mandelbrot1982fractals,bak2013nature,west2000scaling}, urban science~\cite{bettencourt2010unified}, pattern formation~\cite{toffoli1987cellular,schelling2006micromotives}, econophysics~\cite{mantegna1999introduction}, and nonlinear dynamics and chaos~\cite{strogatz2018nonlinear,gleick2011chaos}.  Book series on complex systems topics include the \textit{Santa Fe Institute Series} and \textit{Unifying Themes in Complex Systems}.  

\section{Summary}
\label{sec:conclusion}
Systems with many components often exhibit emergent large-scale behaviors that cannot be directly inferred from the behaviors of their components.  However, an early insight of statistical physics is that in spite of the impossibility of describing the details of trillions of molecules, the macroscopic properties of the molecules can be well understood by analyzing their space of possible behaviors, rather than their specific configurations and motions.  
While many macroscopic properties can be described in terms of the average behaviors of the molecules, the macroscopic properties of certain physical phenomena, such as phase transitions, cannot be understood by averaging over system components; accordingly, physicists were forced to develop new, multi-scale methods.  Likewise, while standard statistical methods---which infer the average properties of a system's many components---can succesfully model some biological and social systems, they fail for others, sometimes spectacularly so. 

Taking a systemic view by considering the space of possible behaviors can yield insights that cannot be gleaned by considering only the proximate causes and effects of particular problems or crises.  A system's complexity---which depends on its number of distinct potential behaviors (i.e. on the space of possibilities)---is a starting point from which to get a handle on its large-scale properties, in the same way that entropy is the starting point for statistical physics.   Because the number of distinct behaviors of a system depends on the level of detail (behaviors that appear the same at lower resolution may be distinct at higher resolution), complexity depends on scale.  Interdependencies between components reduce complexity at smaller scales by restricting the freedom of individual components while creating complexity at larger scales by enabling behaviors that involve multiple components working together.  Thus, for systems that consist of the same components, there is a fundamental tradeoff between the number of behaviors at smaller and larger scales.   This tradeoff among scales is related to the tradeoff between a system's adaptability, which depends on the variety of different responses it has to internal and external disturbances, and its efficiency, which depends on its operating scale.  There is no ideal scale at which a system should possess complexity; rather, the most effective systems are those that at each scale match the complexity of their environments.

When analyzing data or creating organizational structures, standard methods fail when they underestimate the importance of interdependencies and the complexity that arises from these interdependencies.  To some extent, these problems can be mitigated by matching the data analysis or organizational structure to natural divisions within the system of interest.  Since complex systems are those for which behaviors occur over multiple scales, successful organizations and analyses for complex systems must also be multi-scale in nature.  However, even when armed with all the proper information and tools, human understanding of most complex systems will inevitably fall short, with unpredictability being the best prediction.  To confront this reality, we must design systems that are robust to the ignorance of their designers and that, like evolution, are strengthened rather than weakened by unpredictability.  Such systems are flexible with multiple processes occurring in parallel; these processes may compete with one another within a multi-scale cooperative framework such that effective practices are replicated.  Only these systems---that grow in complexity over time from trial and error and the input of many---exhibit the necessary complexity to solve problems that exceed the limits of human comprehension.

\begin{acknowledgments}
This material is based upon work supported by the National Science Foundation Graduate Research Fellowship Program under Grant No. 1122374 and by the Hertz Foundation.  We thank Uyi Stewart for discussions that led to the writing of this paper, Gwendolyn Towers for editing early drafts of the manuscript, and Robi Bhattacharjee for helpful discussions regarding complexity and scale. 
\end{acknowledgments}

\end{document}